\documentclass[12pt,amsmath,amssymb]{article}

\usepackage{graphicx} 
\usepackage{setspace}
\usepackage[margin=1.0in]{geometry}
\usepackage{amsmath}

\usepackage{bm}

\usepackage[mathlines]{lineno}

\def\lteq{\ {\raise-.5ex\hbox{$\buildrel<\over-$}}\ }
\def\apgt{\ {\raise-.5ex\hbox{$\buildrel>\over\sim$}}\ }
\def\aplt{\ {\raise-.5ex\hbox{$\buildrel<\over\sim$}}\ }
\def\lt{\ {\raise-.5ex\hbox{$\buildrel>$}}\ }
\def\gt{\ {\raise-.5ex\hbox{$\buildrel<$}}\ }
\def\eqgt{\ {\raise-.5ex\hbox{$\buildrel>\over-$}}\ }
\def\eqlt{\ {\raise-.5ex\hbox{$\buildrel<\over-$}}\ }

\def\br{{\bf r}}

\def\dch#1{{#1}}

\title{{\bf \Large Punctuated Chaos and Indeterminism in Self-gravitating Many-body Systems}\footnote{Essay written for the Gravity Research Foundation 2023 Awards for Essays on Gravitation.}}

\author{{\bf \normalsize Tjarda C. N. Boekholt}\footnote{e-mail: tjarda.boekholt@physics.ox.ac.uk (corresponding author)}\\ {\normalsize Rudolf Peierls Centre for Theoretical Physics, Clarendon Laboratory,}\\{\normalsize University of Oxford, Parks Road, Oxford OX1 3PU, UK}\\ [3ex] {\bf \normalsize Simon F. Portegies Zwart}\footnote{e-mail: spz@strw.leidenuniv.nl} \\ {\normalsize Leiden Observatory, Leiden University, PO Box 9513, 2300 RA,}\\ {\normalsize Leiden, The Netherlands}\\ [3ex] {\bf \normalsize Douglas C. Heggie}\footnote{e-mail: d.c.heggie@ed.ac.uk} \\ {\normalsize School of Mathematics and Maxwell Institute for Mathematical Sciences,}\\ {\normalsize University of Edinburgh, 10 Kings Buildings, Edinburgh EH9 3FD, UK}\\ [3ex]}

\date{\normalsize \today}
 
\begin{document}

\maketitle

\begin{abstract}
Dynamical chaos is a fundamental manifestation of gravity in astrophysical, many-body systems. 
The spectrum of Lyapunov exponents quantifies the associated exponential response to small perturbations. 
Analytical derivations of these exponents are critical for understanding the stability and predictability of observed systems. 
This essay presents a new model for chaos in systems with eccentric and crossing orbits. 
Here, exponential divergence is not a continuous process but rather the cumulative effect of an ever-increasing linear response driven by discrete events at regular intervals, i.e., punctuated chaos.
We show that long-lived systems with punctuated chaos can magnify Planck length perturbations to astronomical scales within their lifetime, rendering them fundamentally indeterministic.
\end{abstract}

\newpage

\doublespacing

\section*{Measuring chaos in self-gravitating N-body systems}

Newton's gravitational N-body problem \cite{Newton1687} is commonly adopted as a model for the dynamical evolution of astrophysical systems in the weak gravity regime. Applications include the orbital evolution of comets, exoplanets, dense stellar systems, galaxies, and even cosmological structures. Finding analytical solutions for systems with $N>2$ is challenging due to chaos, as pointed out by Poincar\'e \cite{Poincare91, Poincare92}. Over the last few decades, since the advent of the computer, progress has been made from a computational perspective with the development of increasingly fast and accurate N-body algorithms. However, exponential magnification of numerical errors seems unavoidable for chaotic N-body systems, resulting in the calculated trajectory completely diverging from the sought-after mathematical solution to Newton's equations \cite{1964ApJ...140..250M, 2014ApJ...785L...3P}. 
It is an article of faith that such ``approximate'' solutions are nevertheless valid statistically by conserving globally conserved quantities, such as energy and angular momentum \cite{1996IAUS..174..131H}. The statistical mechanics reasoning here is that the numerical N-body system is ergodic, i.e., it explores the available phase space volume completely and in an unbiased manner \cite{2018CNSNS..61..160P}. The reliability of N-body simulations remains a fundamental unsolved problem \cite{2015ComAC...2....2B}. Progress is made by pursuing a better understanding of the growth of small perturbations in chaotic N-body systems. If these perturbations are numerical artifacts, we aim to determine the validity of the ergodic assumption \cite{2014ApJ...785L...3P, TB2020} and construct potential new and improved N-body algorithms \cite{2023MNRAS.519.3281B}. On the other hand, if the perturbations are of a physical nature, then the growth rate of perturbations informs us on the stability and predictability of astrophysical systems \cite{TB2020,2021PhRvD.104h3020B,2022A&A...659A..86P,2022arXiv220903347P}. For all of these various reasons, being able to resolve and accurately measure the growth of small perturbations is essential.    

To study chaos in N-body systems, we developed the most accurate and precise N-body code to date, called \texttt{Brutus} \cite{2015ComAC...2....2B}. We achieve this feature by solving Newton's equations of motion by brute force. Firstly, we implement the Bulirsch-Stoer integration method, which consists of an iterative integration scheme with extrapolation to zero step size. Secondly, we replace the conventional double-precision arithmetic with an arbitrary-precision arithmetic software library. The latter allows us to define the number of decimal places to represent a number, only limited by the computer's memory. Using \texttt{Brutus} we control the discretisation and round-off errors and systematically reduce their magnitude to the point of numerical convergence. We define the converged solution as a numerical solution for which the first specified number of decimal places \dch{have converged, and assume that these decimal places are then} 
the same as \dch{in} the mathematical solution \cite{2018CNSNS..61..160P}. Hence, the main novelty of our experimental approach is that we can study dynamical chaos with converged solutions to the N-body problem \cite{2015ComAC...2....2B, 2016MNRAS.461.3576B, 2018CNSNS..61..160P, TB2020, 2021PhRvD.104h3020B}. Numerically converged solutions are crucial for distinguishing between physical chaos and numerical noise. 

Various methods can be found in the literature for measuring the growth rates of small perturbations. These are typically expressed in terms of Lyapunov exponents or time scales. One method is to integrate Newton's equations of motion and the corresponding variational equations. However, if the solution is not converged, this variational method measures the Lyapunov exponents of an ``approximate'' solution, which thus relies on the validity of the ergodic assumption. Alternatively, we consider \dch{the difference between a solution using} the initial conditions of interest and \dch{another which uses slightly altered initial conditions.} 
We integrate these neighboring trajectories up to convergence using \texttt{Brutus} \cite{2015ComAC...2....2B}. This results in a converged solution for the perturbation growth and the associated measures of the Lyapunov exponent\dch{, provided that the difference between the two solutions remains small}.    

The origin of chaos and its associated exponential sensitivity depends on the configuration of the N-body system. For planetary systems, in which orbits are typically non-overlapping (circular and planar), chaos is mostly driven by an overlap of orbital resonances. This does not apply to many other astrophysical systems, which consist of overlapping (eccentric and inclined) orbits. In these cases, chaos is thought to be driven by many random close encounters, somewhat resembling a violent game of pool. A dense stellar system consisting of millions of equal-mass bodies is indeed found to be chaotic with a Lyapunov time scale of only a fraction of its dynamical time \cite{1993ApJ...415..715G,2022A&A...659A..86P}. 
We aim to use \texttt{Brutus} to measure the exponential sensitivity of chaotic N-body systems accurately and to correlate it with the orbital dynamics. 

\begin{figure}
\centering
\begin{tabular}{c}
\includegraphics[height=0.6533\textwidth,width=0.98\textwidth]{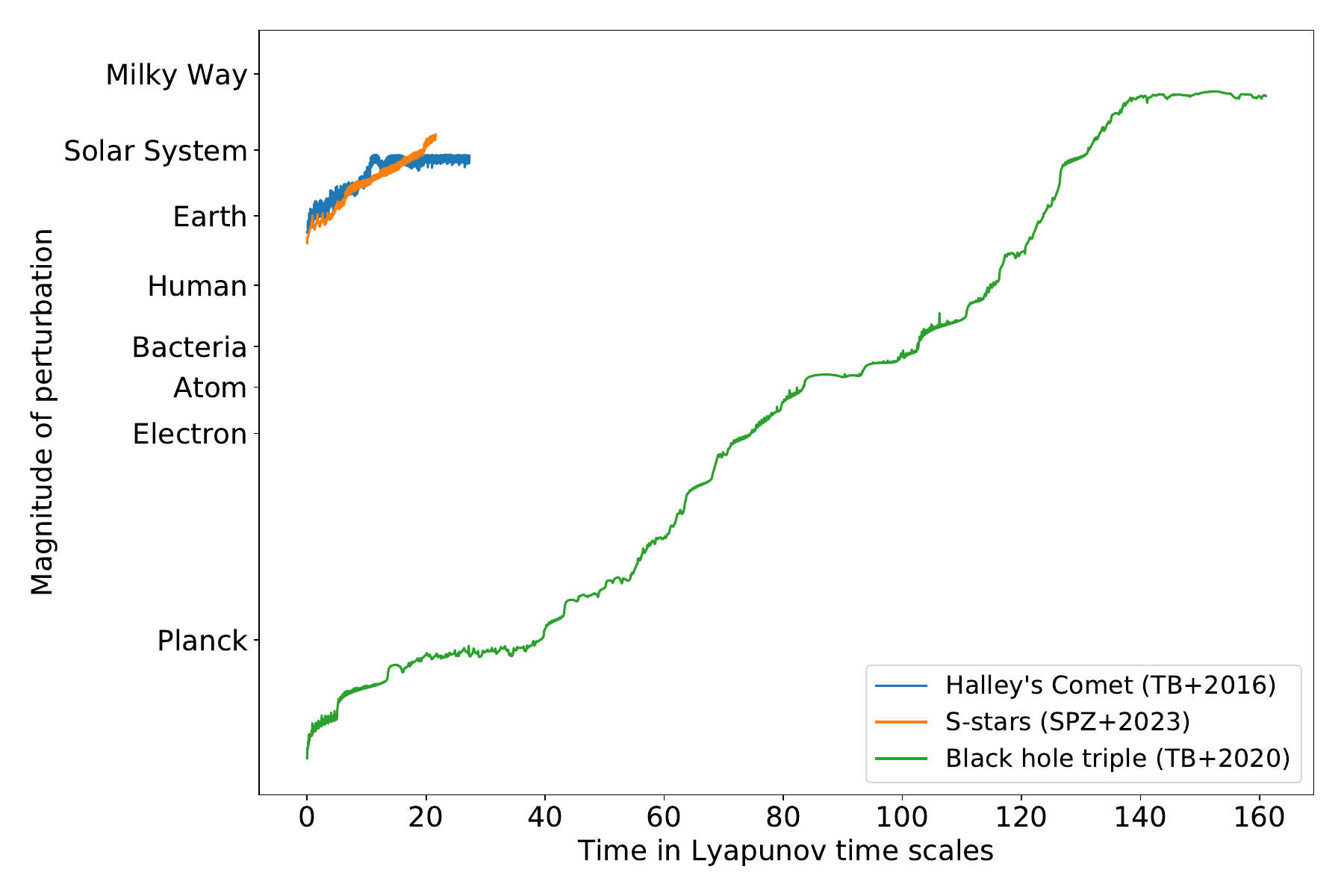} \\
\end{tabular}  
\caption{ Measuring the exponential sensitivity of three different astrophysical systems. We plot the time evolution of an initially small perturbation to the system. Time is normalised by the Lyapunov time. For the black hole triple system, we observe that an initial quantum fluctuation is magnified exponentially to astronomical scales. Such fundamentally indeterministic systems are a common feature of chaotic astrophysical populations \cite{TB_inprep_agekyan_with_L}.  }
\label{fig:1}
\end{figure}

We apply our novel \texttt{Brutus} method to obtain converged solutions to three different N-body problems. We consider: 1) Halley's Comet and its interaction with the Sun and planets \cite{2016MNRAS.461.3576B}, 2) the population of S-stars in our Galactic center, which closely orbit the supermassive black hole \cite{SPZ_inprep}, and 3) chaotic triple systems consisting of supermassive black holes \cite{TB2020}. Each of these configurations is chaotic and consists of crossing orbits. 
In Fig.~\ref{fig:1}, we plot the magnitude of a small initial \dch{variation} 
and its detailed growth in time. First, we note that initial values for perturbations are typically set to the uncertainty of the observations, which range from meters to thousands of kilometers for Solar System bodies. We adopted this approach for the case 
of Halley's Comet. 
\dch{This} 
system 
exponentially magnif\dch{ies} 
this initial uncertainty to the size of the system itself within a mere 10-20 Lyapunov time scales\dch{, after which the difference between the two solutions is no longer small.  In such a case} an improved measure of the Lyapunov exponent can be obtained if we reduce the initial magnitude of the perturbation, allowing us to measure the time-averaged exponential growth rate over many more Lyapunov time scales \dch{(if necessary)}. This approach was adopted for the  triple system of supermassive black holes illustrated in Fig.~\ref{fig:1}. In doing so, we discovered solutions where an initial perturbation of order of the Planck length (and smaller) is exponentially magnified to astronomical scales over 100+ Lyapunov time scales, at which point the interaction among the black holes is still ongoing. This example has profound implications on the role of chaos as it directly connects Heisenberg's quantum uncertainty principle with the evolution of self-gravitating systems on astronomical scales \cite{2022arXiv220903347P}. Three bodies are sufficient to introduce an arrow of time \cite{TB2020}  

A closer inspection of the results in Fig.~\ref{fig:1} reveals that, on average, a well-defined slope exists\dch{, at least until the variation becomes too large, as in the Halley example}. The time-averaged Lyapunov exponent is thus a well-defined characteristic of a chaotic N-body system. Superposed, however, there are stochastic fluctuations and jumps.  \dch{As we discuss later, in connection with Fig.~\ref{fig:2},} a recurring feature in the curve is a sudden strong jump followed by a gradual flattening 
consistent with linear growth. The overall growth can then be interpreted as \dch{the} accumulati\dch{on of} 
\dch{successive periods of}  
linear response driven by events that occur at discrete moments. This leads to the idea of a ``{\it punctuated chaos}'' as opposed to a smooth continuous exponential growth \cite{SPZ_inprep}   

\section*{Theory of Punctuated Chaos}

The above observations inspired us to develop a new theoretical model for chaos in self-gravitating N-body systems. This model captures the punctuated growth of perturbations and relates the moments of abrupt change in the growth to events occurring in the N-body system.
By relating the detailed orbital dynamics to the punctuated growth of perturbations, we ultimately obtain a theoretical expression for the time-averaged Lyapunov exponent. 

The derivation of the model of punctuated chaos starts by considering the case of a body in a Kepler orbit around a much more massive body (as is the case for Halley's Comet and the S-stars).
With an initial semi-major axis $a_0$ and total mass $m$, the initial
orbital frequency $\omega_0 = \sqrt{Gm/a^3_0}$. Let a neighboring solution
be separated by an infinitesimal displacement $\delta x_0$ at some
time $t=0$. This displacement has components along and
transverse to the orbit. The displacement leads to a small difference
in the semi-major axis of the same order, i.e., $\delta a_0 \sim \delta x_0$.
The resulting difference in frequency is
\begin{equation}
\delta \omega_0 \sim \delta x_0 \sqrt{Gm/a^5}. 
\label{Eq:delta_orbital_frequency}
\end{equation}
The displacement along the orbit grows with time $t>0$ according to
\begin{equation}
\delta x(t) \sim \delta x_0 + \delta \omega_0 a_0 t = \delta x_0 (1 + \omega_0 t).
\label{Eq:delta_xt}\end{equation}
The growth of the initial displacement is linear with time from $t_0$
to $t$, but such that $\delta a$ remains constant as the growth is along
the orbit, i.e., the growth is in orbital phase rather than in energy and angular momentum. 

Now suppose that an instantaneous perturbation acts on the motion at time
$t_1$, causing the velocity of the Kepler motion to receive a slight kick.
\dch{We suppose that the variation in semi-major axis is again of order the spatial variation, i.e.} 
$\delta a_1 \sim \delta x_1$, and
this leads to a difference in orbital frequency $\delta \omega_1 \sim
\delta a_1 \omega_1 /a_1$ at time $t_1$. Thus for $t > t_1$ the
displacement varies as
\begin{equation}
\delta x \sim \delta x_1 + \omega_1 (t - t_1)
\delta a_1 \sim \delta x_1 (1 + \omega_1 (t - t_1)).
\label{Eq:delta_x_and_a}\end{equation}
If a second kick occurs at time $t_2 > t_1$, we can see from
Eqs.~\eqref{Eq:delta_xt} and \eqref{Eq:delta_x_and_a} that the displacement
is
\begin{equation}
\delta x_2 \sim \delta x_0 (1 + \omega_0 t_1)(1 + \omega_1 (t_2 - t_1)).
\label{Eq:delta_x2}\end{equation}

\noindent If these perturbations recur at roughly comparable intervals $\Delta
t$, and if $\omega$ does not change by a large factor, it can be seen
that the displacement at some large time $t$ will be
\begin{equation}
\delta x(t) \sim \delta x_0 (1 + \omega \Delta t)^{t/\Delta t}.
\end{equation}

\noindent The linear growth of Eq.~\eqref{Eq:delta_xt} transforms into
exponential growth! The corresponding Lyapunov exponent is ${\cal
O}(\omega)$ if $\omega \Delta t \aplt 1$; it is of order the
reciprocal of the dynamical time (also called ``crossing time''). The case $\omega \Delta t \apgt 1$
is also of interest and leads to a smaller estimate of order
$\ln(\omega \Delta t)/\Delta t$. Hence, the time-averaged Lyapunov exponent is determined by the frequency 
of events.  \dch{Furthermore, though we have ignored this issue so far, it is also affected by their magnitude.}

\begin{figure}
\centering
\begin{tabular}{c}
\includegraphics[height=0.784\textwidth,width=0.98\textwidth]{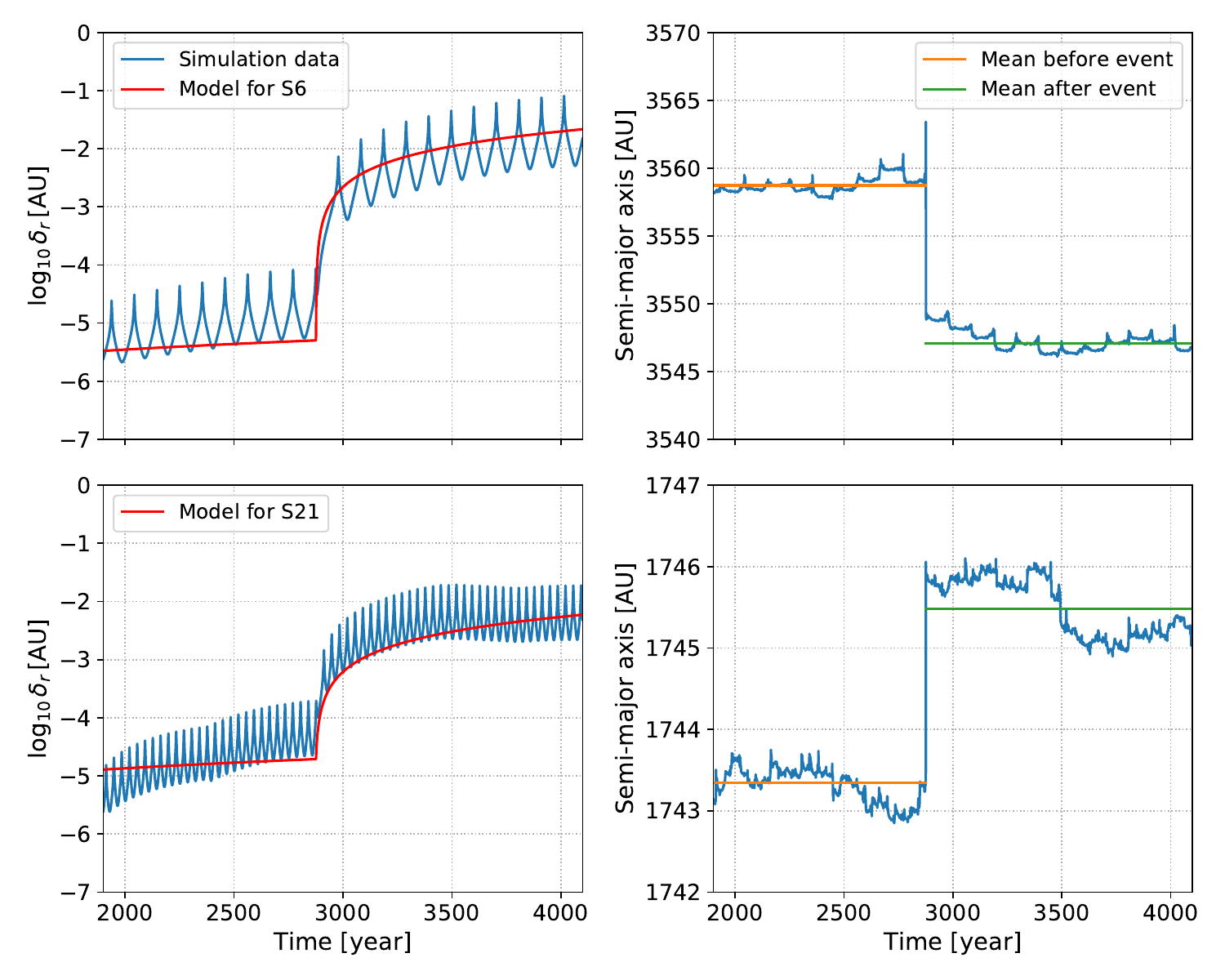} \\
\end{tabular}  
\caption{ A demonstration of punctuated chaos. Two stars called S6 and S21 engage in a very brief but strong encounter. This results in an approximately instantaneous energy exchange. Their orbital elements, such as semi-major axis (right column) change in a step-like manner. The magnitude of \dch{the variation} 
  (left column) displays the characteristic \dch{linear} 
  growth driven by the close encounter\dch{, here plotted logarithmically}. The cumulative effect of multiple such \dch{events} 
  results in net exponential growth. }
\label{fig:2}
\end{figure}

To determine the trigger of an event, we inspect the orbital dynamics around the times when an event is detected. We focus on the S-star configuration, and in Fig.~\ref{fig:2}, we zoom in on the perturbation growth around a single event occurring just before $t=3000\,\rm{years}$. We observe that the magnitude of the \dch{variation in energy} 
is abruptly magnified\dch{,} in accordance with a punctuated event. The superposed shorter-term oscillations are due to the eccentricity of the orbit. The same figure correlates the event with a close encounter between stars S6 and S21. These two stars exchange orbital energy causing S6 to abruptly shrink its orbit, while S21 expands its orbit proportionally. This example illustrates the underlying mechanism where events are generally triggered by strong few-body interactions.    

In order to estimate the effect of a close encounter on the perturbation growth, we will reduce the problem to a 3-body system consisting of a black hole (mass $M$, particle 0) and two S-stars (mass $m$, particles 1,2).  Let $\br_i$ be the position vector of star $i$ relative to the black hole, and focus on star 1.  Its equation of motion is
\begin{eqnarray}
  \ddot{\br}_1 &=& -\frac{G(M+m)\br_1}{r_{1}^3} - \frac{Gm(\br_1 - \br_2)}{r_{12}^3} - \frac{Gm\br_2}{r_2^3}\label{eq:eom}
\end{eqnarray}
with obvious notation $r_i,r_{12}$.   The first term on the right is the acceleration of star 1  relative to the black hole, the second term is the direct perturbation by particle 2, and the third term is the indirect perturbation, caused by the acceleration of the black hole by particle 2. 
After inspecting the magnitudes of each of these components, we find that the perturbation on the motion of particle 1 is dominated by the second term on the right of Eq.~\eqref{eq:eom}, i.e. the direct perturbation due to the gravitational attraction of star 2. 

We now consider the growth of a perturbation in the single S-star's orbit which we have hitherto labeled as star 1, and suppose that it has a succession of close encounters with other stars (in ``events'') at some interval $\Delta t$.  Let $\delta x$ denote the variation in some quantity $x$, such as energy or position, between two neighbouring solutions.  Specifically, let $\delta E_0$ be the variation in specific energy $E\equiv\vert v^2/2 - GM/r\vert$ at a time just before an event, where $v$ is the speed, and let $\delta r_0$ be the variation in position at the same time.  (Henceforth the subscripts denote an index of events in a sequence of events.)

Then we have
\begin{equation}
  \delta E_1 = \delta E_0 + \delta\Delta E,\label{eq:DeltaE}
\end{equation}
where $\delta E_1$ is the variation just before the next event, which will be the same as the value just {\sl after} the first event; and $\Delta E$ is the change in $E$ at that first event.  Now we estimate this as
\begin{equation}
  \Delta E = v.\Delta v \sim \sqrt{GM/r}\frac{Gm}{dV},
\end{equation}
where 
$d$ is the distance of the closest approach in the encounter, and $V$ is the relative speed of the two stars in the encounter. (This estimate is made by taking $\Delta v$ to be the maximum perturbation $Gm/d^2$ \dch{multiplied} by the time scale of the encounter, $d/V$.)  We use the same estimate $\sqrt{GM/r}$ for $V$ as for $v$, so that
\begin{equation}
\Delta E\sim Gm/d,
\end{equation}
Its variation can  be estimated as 
\begin{equation}
\delta(\Delta E) \sim \frac{Gm}{d^2}\delta r_0,
\end{equation}
and so Eq.~\eqref{eq:DeltaE} becomes
\begin{equation}
\delta E_1 = \delta E_0 + \frac{Gm}{d^2}\delta r_0.\label{eq:delta-E1}
\end{equation}

\noindent After this event the variation in the orbital frequency will be $\delta\omega_1$, and so
\begin{equation}
  \delta r_1 \sim \delta r_0 + \delta\omega_1 a \Delta t,
\end{equation}
where $a$ is the semi-major axis. Since $\omega^2a^3 = GM$ and $E \sim GM/a$, we can re-express this as
\begin{equation}
    \delta r_1 \sim \delta r_0 + \frac{\delta E_1\Delta t}{\sqrt{\frac{GM}{a}}}.
\end{equation}
By Eq.~\eqref{eq:delta-E1}, this in turn becomes
\begin{equation}
      \delta r_1 \sim \delta r_0 + \frac{\left(\delta E_0 + \frac{Gm}{d^2}\delta r_0 \right)\Delta t}{\sqrt{\frac{GM}{a}}}.\label{eq:delta-r1}
\end{equation}

\noindent Equations \eqref{eq:delta-E1} and \eqref{eq:delta-r1} are explicit estimates which allow us to map the effect of an encounter on $\delta E$ and $\delta r$.  It has matrix
\begin{equation}
  A = 
  \left(\begin{array}{cc}
    1&\frac{Gm}{d^2}\\
    \Delta t/\sqrt{GM/a}&1 + \frac{Gm}{d^2}\Delta t/\sqrt{GM/a}
  \end{array}\right).
\end{equation}
The largest eigenvalue of this matrix gives the factor by which the variation, in either $E$ or $r$, is multiplied as a result of an event.

The analysis above focused on the direct perturbation between two stars, i.e. a close encounter. This encounter triggered the event, which led to punctuation in the exponential sensitivity of the entire N-body system. 
The mechanism which propagates the effect of a few-body encounter to the rest of the N-body system is the indirect perturbation described by the third term on the right in Eq.~\eqref{eq:eom}. 
It can be shown that the response of a star not involved in the event follows the same Lyapunov exponent as the stars involved \cite{SPZ_inprep}. The time-averaged Lyapunov exponent is thus shared among all bodies \dch{through the mediation of the central black hole}, and therefore \dch{becomes} a fundamental property of the N-body system as a whole.

To conclude, in this essay we motivated the importance of pursuing a better understanding of the growth of perturbations in self-gravitating N-body systems. To this end, we developed the unique N-body code \texttt{Brutus}, which allows us to study chaos with converged solutions. Accurate measurements of chaos in various astrophysical systems led to the development of a new theory of chaos: ``punctuated chaos''. Here, exponential divergence is the net result of a discrete set of events, i.e. strong few-body encounters, which each trigger a 
linear response. If the lifetime of a chaotic N-body interaction exceeds 100+ Lyapunov time scales, we demonstrated that punctuated chaos can connect the fundamentally uncertain quantum world to the dynamical evolution of self-gravitating N-body systems on astronomical scales.  


\bibliographystyle{plain}
\bibliography{grf} 

\end{document}